\documentclass{ws-procs10x7}
\usepackage{epsfig}

\newcommand{\lsim}{\raisebox{-0.13cm}{~\shortstack{$<$ \\[-0.07cm] $\sim$}}~}
\newcommand{\gsim}{\raisebox{-0.13cm}{~\shortstack{$>$ \\[-0.07cm] $\sim$}}~}

\newcommand{\ee}{$e^+e^-$}

\newcommand{\beq}{\begin{eqnarray}}
\newcommand{\eeq}{\end{eqnarray}}

\catcode`@=11
\def\citer{\@ifnextchar
[{\@tempswatrue\@citexr}{\@tempswafalse\@citexr[]}}

\def\@citexr[#1]#2{\if@filesw\immediate\write\@auxout{\string\citation{#2}}\fi
  \def\@citea{}\@cite{\@for\@citeb:=#2\do
    {\@citea\def\@citea{--\penalty\@m}\@ifundefined
       {b@\@citeb}{{\bf ?}\@warning
       {Citation `\@citeb' on page \thepage \space undefined}}%
\hbox{\csname b@\@citeb\endcsname}}}{#1}}
\catcode`@=12

\begin{document}
\title{Linear collider prospects on electroweak physics\footnotemark[1]}

\author{Margarete M\"uhlleitner}

\address{Paul Scherrer Institut, CH-5232 Villigen PSI, Switzerland}

\twocolumn[\maketitle\abstract{
Prospects on electroweak physics at a future International Linear Collider 
(ILC) are summarized, including gauge coupling measurements, 
top quark physics and Higgs physics. 
\begin{center}
\centerline{\phantom{x}\hspace*{-0.8cm} hep-ph/0411059 
\hspace*{1cm} PSI-PR-04-13}
\end{center}
\vspace*{-0.2cm}
}]

\footnotetext{$^*$Invited talk at the 32nd International Conference on High 
Energy Physics, ICHEP04, 16-22 August 2004, Beijing, China.}

\section{Introduction}
\vspace*{-0.2cm}
A major goal of present and future research in high-energy physics is
the understanding of the mechanism of mass generation and electroweak 
symmetry breaking (EWSB). Precision measurements in the electroweak sector 
of the Standard Model (SM) and its extensions allow us to tackle these
questions both directly and indirectly via quantum effects. The clean 
environment of $e^+e^-$ colliders\cite{ilc} with high luminosity sets the 
basis for approaching this task. In the following I will exemplify with the 
help of some prominent results the prospects for electroweak measurements at 
future \ee colliders in the gauge boson, top quark and Higgs boson sector.


\vspace*{-0.3cm}
\section{Electroweak gauge bosons}
\vspace*{-0.2cm}
A primary goal for the study of gauge boson properties is to establish 
the non-Abelian nature of electroweak interactions. Very precise measurements
constrain new physics at scales above the direct reach of the 
machine. Processes sensitive to triple gauge
couplings in \ee collisions are $W$ production in pairs, $e^+e^-\to W^+W^-$, 
or singly in $e^+e^-\to We\nu$. At high luminosity and with the help of 
beam polarisation the triple couplings can be determined with an error of
a few $10^{-4}$, see Table~\ref{tab1}\cite{menges}, so that new physics at
high scales can be tested.

In case there is no light Higgs particle unitarity requires the gauge bosons
to become strongly interacting at $\sqrt{s}\gsim 1$~TeV. Anomalous quartic
couplings can be probed in gauge boson scattering\cite{kilian} where six 
fermion 
final states have to be studied. Simulations have shown that at an \ee 
collider electroweak symmetry breaking scales up to 3~TeV can be probed\cite{rosati} covering the threshold region of strong $WW$ interactions.  
\begin{table}
\begin{center}
{\small
\begin{tabular}{|c|c|c|}
\hline
coupling & \multicolumn{2}{|c|}{error $\times 10^{-4}$} \\
\cline{2-3}
& $\sqrt{s}=500$~GeV & $\sqrt{s}=800$~GeV \\
\hline
\multicolumn{3}{|c|}{C,P-conserving, $SU(2)\times U(1)$ relations:} \\
\hline
$\Delta g_1^Z$ & 2.8 & 1.8 \\
$\Delta \kappa_\gamma$ & 3.1 & 1.9 \\
$\lambda_\gamma$ & 4.3 & 2.6 \\
\hline
\multicolumn{3}{|c|}{C,P-conserving, no relations:} \\
\hline
$\Delta g_1^Z$ & 15.5 & 12.6 \\
$\Delta \kappa_\gamma$ & 3.3 & 1.9 \\
$\lambda_\gamma$ & 5.9 & 3.3 \\
$\Delta\kappa_Z$ & 3.2 & 1.9 \\
$\lambda_Z$ & 6.7 & 3.0 \\
\hline
\multicolumn{3}{|c|}{not C or P conserving} \\
\hline
$g_5^Z$ & 16.5 & 14.4 \\
$g_4^Z$ & 45.9 & 18.3 \\
$\tilde\kappa_Z$ & 39.0 & 14.3 \\
$\tilde\lambda_Z$ & 7.5 & 3.0 \\
\hline
\end{tabular}
}
\end{center}
\caption{Single parameters fits ($1\sigma$) to triple gauge couplings with 
beam polarisation ${\cal P}_{e^-/e^+}=80/60$\%. Parametrisation as in 
Ref.$^3$.}
\label{tab1}
\vspace*{-0.5cm}
\end{table}


\vspace*{-0.3cm}
\section{Top quark physics}
\vspace*{-0.2cm}
The top quark with a mass $m_t=178\pm 4.3$~GeV\cite{top} is the heaviest 
observed fermion. With its lifetime being much larger than the QCD scale top 
production and decay can be analysed within perturbative QCD. New interactions 
may be revealed through non-standard top decays. 

\begin{figure}[ht!]
\begin{center}
\epsfig{figure=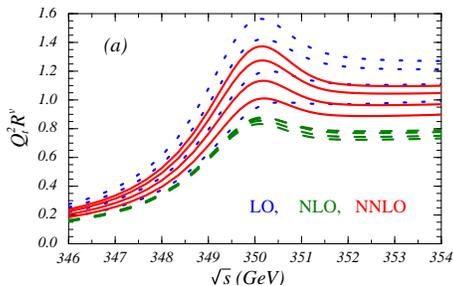,bbllx=90,bblly=440,bburx=532,bbury=715,width=6cm,clip=}
\end{center}
\caption{$Q_t^2 R^v$ with fixed $M_t^{1S}$ for the resummed expansion. 
See Ref.$^{9}$.}
\vspace*{-0.3cm}
\label{topfig}
\end{figure}

The top quark mass can be precisely measured in threshold production. On
the theoretical side a lot of progress in determining the
threshold cross section has been made: Threshold masses have been 
introduced\cite{thresh} to stabilize the location of the threshold and to 
reduce the correlation between $m_t$ and the strong coupling constant 
$\alpha_s$. The overall normalization of the 
cross section has been improved by the introduction of renormalization group
improved perturbation theory where large QCD logarithms are 
resummed\cite{resum}, see Fig.\ref{topfig}.
A full NNLL order prediction, though almost complete, is still 
missing. The present estimate on the cross section is still under discussion 
and of order $\pm 6$\%\cite{hoang}. In
the alternative fixed order perturbation series important progress has been 
made for the NNNLO contributions\cite{penin}. 

On the experimental side an updated $t\bar{t}$ threshold scan simulation has
been performed\cite{martinez}. It includes not only experimental systematic
errors but also an estimate of the theoretical error in the cross section
prediction. By performing a multiparameter fit it takes 
into account the large correlations between the physical parameters.
The top mass, the top width and $\alpha_s(M_Z)$ can
be extracted simultaneously with uncertainties of about 20 MeV, 30 MeV and
0.0012, respectively. The extraction of the top Yukawa coupling from a 
four parameter fit, however, suffers from an error
of several tens of percent. The current theoretical error of about 100 MeV
on the top quark mass has not been included.

Anomalous top quark couplings can be probed in continuum production 
at high energies\cite{breuther}.

\vspace*{-0.3cm}
\section{Higgs physics}
\vspace*{-0.2cm}
The Higgs mechanism is a cornerstone in the electroweak sector 
of the SM and its supersymmetric (SUSY) extensions. It allows to generate
particle masses without violating gauge principles. In order to establish
the Higgs mechanism experimentally four steps have to be taken: the Higgs
particle(s) must be discovered, the spin and CP properties have to be 
determined, the gauge and Yukawa couplings must be measured and finally the
Higgs self-interactions are to be determined to reconstruct the Higgs potential
itself. 

The main SM Higgs boson production processes are Higgs-strahlung\cite{hrad}
at low energies, $e^+e^-\!\to\! Zh$, and $WW$ fusion\cite{fusion} at high 
energies, $e^+e^- \!\to\! H\nu\bar\nu$. The 
full electroweak (EW) corrections at one loop have been calculated
for both the Higgs-strahlung\cite{nlohrad,nloboth} and the fusion 
process\cite{nloboth,nlowfus}. They are of ${\cal O}(10\%)$. 
By combining recoil mass techniques and reconstruction of the Higgs decay 
products, the accuracy on $M_H$ is 40-80 MeV for intermediate Higgs 
bosons\cite{garcia}. Furthermore, the Higgs boson couplings to massive gauge 
bosons are best probed in the two production processes. 
The accuracies on the total cross sections\cite{hhz} and branching 
ratios\cite{branch} are summarized in Table~2.
\begin{table}[h]
\begin{center}
{\small
\begin{tabular}{|l|l|}
\hline
$(\delta\sigma/\sigma)_{ZH}\!$ & 2.5...3~\% 
\\
$(\delta BR/BR)_{ZZ}\!$ & 17~\% 
\\
\cline{1-2}
$(\delta\sigma/\sigma)_{WW}\!$ & 2.8...13\%
\\
$(\delta BR/BR)_{WW}\!$ & 5.1...2.1~\% 
\\
\hline
\end{tabular}
}
\caption{Accuracies on SM Higgs boson production cross sections 
and branching ratios into $WW/ZZ$ for $M_H=120$-$160$~GeV (160 GeV for 
$BR(H\to ZZ)$).}
\end{center}
\label{table4}
\vspace*{-0.3cm}
\end{table}

\begin{figure}[h]
\begin{center}
\epsfig{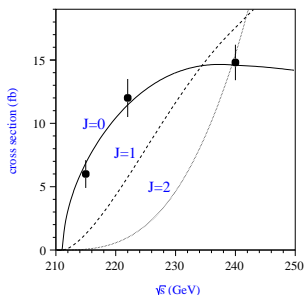}
\end{center}
\caption{The threshold cross section $e^+e^- \!\to\! ZH \to l^+ l^- + 2$~jets
at three c.m. energies and the predictions for spin s=0 (full line), s=1 
(dashed line) and s=2 (dotted line).}
\vspace*{-0.3cm}
\label{hzfig}
\end{figure}

The spin and CP properties can be determined in a model-independent way 
from the angular distribution of the $Z$ boson in 
$e^+e^-\!\to\! ZH$\cite{anghrad}. Another method exploits the 
threshold dependence of the excitation curve together with the angular 
distribution\cite{djmiller}. An experimental study\cite{hz} shows that 
already with $\int{\cal L}=20$~fb$^{-1}$ the measurement of the 
threshold cross section at three c.m. energies allows the confirmation of the 
scalar nature of the Higgs bosons, see Fig.\ref{hzfig}. For $M_H\!<\!2 M_Z$ 
the spin can also be determined from the invariant mass spectrum in the decay 
$H\!\to\! ZZ^*$ supplemented by angular correlations\cite{choi}.

The Higgs branching ratios into fermions can be measured with a precision at
the percent level at a future ILC\cite{battaglia} by combining the 
measurements of the total Higgs production cross sections with 
$\sigma_{HZ,H\nu\bar{\nu}}\times BR(H\!\to\! f\bar{f})$. Another 
method\cite{brient} determines the fraction of $H\!\to\! X$ decay events in a
sample of unbiased $HZ$ events. 

\begin{figure}[h]
\begin{center}
\epsfig{figure=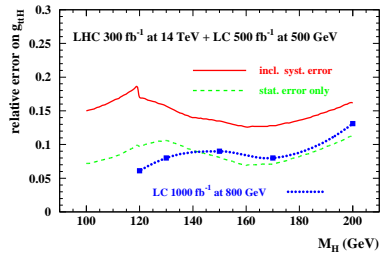,bbllx=0,bblly=0,bburx=500,bbury=334,width=5cm,clip=}
\end{center}
\caption{Precision on the top Higgs Yukawa coupling taking LHC and LC data. For
comparison the precision for the LC alone is also shown.}
\label{figtth}
\vspace*{-0.3cm}
\end{figure}

The Higgs top Yukawa coupling is best measured in 
$e^+e^- \!\to\! t\bar{t} H$\cite{tth} for $M_H\!<\! 2 m_t$. A new experimental 
study\cite{agay} reports expected top Yukawa coupling uncertainties of 6-14\% 
for $120\!<\!M_H\!<\!200$~GeV with $\int{\cal L}\!=\!1$~ab$^{-1}$, 
$\sqrt{s}\!=\!800$~GeV. Taking advantage of a possible synergy of LHC and ILC, 
Ref.\cite{deschtth} gives a 15\% accuraccy in the same mass range, see 
Fig.\ref{figtth}.

For $M_H\!>\!2 m_t$, the Higgs top Yukawa coupling is determined from the 
branching ratio $BR(H\!\to\! t\bar{t})$. The expected experimental accuracy is 
5(12)\% for $M_H\! =\! 400(500)$~GeV at $\int {\cal L} \!=\! 1$~ab$^{-1}$, 
$\sqrt{s} \!=\! 800$~GeV\cite{alcaraz}.

The absolute values of the Higgs couplings are extracted from a global fit to 
the measurable observables, {\it i.e.}~the production cross sections and 
measured branching ratios discussed above. 
This is the only method by which the couplings 
can be determined in a model-independent way. A program 
{\tt HFITTER}\cite{hfitter} has been developed for that purpose. 
Table~\ref{couplgs} shows the achievable accuracies for the couplings. The 
coupling measurement serves as a first crucial test for the Higgs mechanism 
which predicts the couplings to be proportional to the mass of the respective 
particle.
\begin{table}[h]
\begin{center}
{\small
\begin{tabular}{|lll|}
\hline
Coupling & $M_H=120$~GeV & 140~GeV \\
\hline
$g_{HWW}$ & $\pm 0.012$ & $\pm 0.020$ \\
$g_{HZZ}$ & $\pm 0.012$ & $\pm 0.013$ \\
\hline
$g_{Htt}$ & $\pm 0.030$ & $\pm 0.061$ \\
$g_{Hbb}$ & $\pm 0.022$ & $\pm 0.022$ \\
$g_{Hcc}$ & $\pm 0.037$ & $\pm 0.102$ \\
\hline
$g_{H\tau\tau}$ & $\pm 0.033$ & $\pm 0.048$ \\
\hline
\end{tabular}
}
\end{center}
\caption{Relative accuracy on the Higgs couplings assuming $\int\!{\cal L}\!=\!500$~fb$^{-1}$, $\sqrt{s}\!=\!500$~GeV ($\int\!{\cal L}\!=\!1$~ab$^{-1}$, $\sqrt{s}\!=\!800$~GeV 
for $g_{Htt}$).}
\label{couplgs}
\vspace*{-0.4cm}
\end{table}

The lifetime $\Gamma_H$ of the Higgs, being rather small for 
$M_H\lsim 200$~GeV, can be extracted indirectly by combining coupling with 
branching ratio measurements. In the $WW$ channel accuracies of 4-13\% for
$M_H=120$-$160$~GeV can be reached\cite{hhz,branch}.

In the SM the trilinear and quartic Higgs self-couplings are uniquely 
determined by the mass of the Higgs particle. The measurement of 
$\lambda_{HHH}$ hence serves as a consistency check of the SM Higgs mechanism. 
At the ILC it is accessible\cite{3lam} in double Higgs-strahlung 
$e^+e^-\!\to\! ZHH$\cite{hhzproc} at low energies and in $WW$ fusion into 
Higgs pairs at high energies, $e^+e^- \!\to\! HH\nu\bar\nu$\cite{wwhh}. Since 
the cross sections of only a few fb are rather small the highest possible 
luminosities are needed. Experimental studies have shown that 
$\lambda_{HHH}$ can be extracted from $e^+e^-\!\to\! ZHH$ with better than 
20\% for $M_H\!=\!120$~GeV and $\sqrt{s}\!=\!500$~GeV, 
$\int{\cal L}\!=\!1$~ab$^{-1\;}$\cite{castanier}. At a multi-TeV collider the 
expected error is about 8\% for $M_H=120$-$180$~GeV\cite{boos}. A further
recent study reports a possible 10\% measurement by exploiting $WW$ fusion and 
Higgs-strahlung\cite{yamashita}.

The Higgs sector of the Minimal Supersymmetric Extension of the SM (MSSM)
consists of 5 Higgs particles, 2 CP-even, $h,H$, one CP-odd, $A$, and 
two charged ones, $H^\pm$. The heavy Higgs particles can be produced in 
\ee collisions in pairs, $e^+e^-\to HA$. A recent experimental
study has shown, that the $H,A$ masses can be measured with a several
hundred MeV accuracy for Higgs pair production far above the kinematic
threshold\cite{raspereza}. Charged Higgs bosons with $M_{H^\pm}\!<\!\sqrt{s}/2$
can be pair produced. The expected mass resolution for $M_{H^\pm}\!=\!300$~GeV
is 1.5\%\cite{ferrari}. 

\begin{figure}[h]
\begin{center}
\epsfig{figure=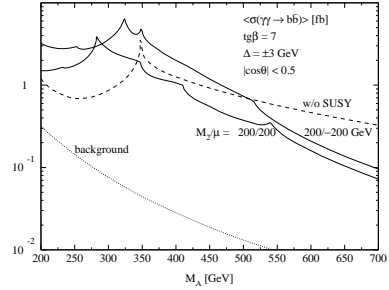,bbllx=49,bblly=222,bburx=565,bbury=618,width=5cm,clip=}
\end{center}
\caption{Cross section for resonant $H,A$ production in $\gamma\gamma$ 
collisions with final decays into $b\bar{b}$ and the corresponding background 
cross section for different MSSM parameters.}
\label{gagafus}
\vspace*{-0.3cm}
\end{figure}

Furthermore, heavy MSSM Higgs bosons can be produced as $s$ channel 
resonances in photon collisions\cite{plc} for  $M_{H/A}\gsim 200$~GeV and
medium values of $\tan\beta$, a parameter region which is not accessible in the
\ee mode for masses above $\sqrt{s}/2$ and in which the LHC might be blind
for the $H/A$ discovery. A simulation of the $b\bar{b}$ final state\cite{gg} 
(see Fig.\ref{gagafus}), finds that the cross section can be determined with
a statistical precision of of 8-20\%\cite{niez}.

\begin{figure}[h]
\begin{center}
\epsfig{figure=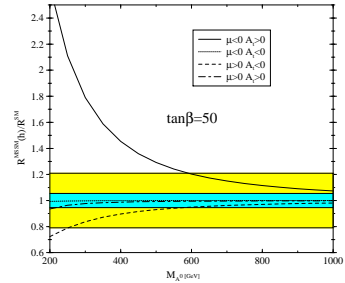,bbllx=52,bblly=51,bburx=523,bbury=450,width=4.4cm,clip=}
\end{center}
\caption{Deviation of $R^{MSSM}(h)$ with respect to the SM value as a 
function of $M_A$ for different SUSY parameters.}
\label{figguasch}
\vspace*{-0.3cm}
\end{figure}

The MSSM Higgs boson couplings are modified with respect to the SM couplings 
so that a precise determination of the couplings may distinguish the two 
models. A measurement of the ratio $BR(h\!\to\! b\bar{b})/BR(h\!\to\! WW^*)$ 
gives indirect access to $M_A$\cite{battaglia}. Combining LHC and LC data 
an accuracy of 20\% (30\%) for $M_A=600\,(800)$~GeV seems to be 
feasible\cite{gross}. A precise determination of 
$R(h)\!=\!BR(h\!\to\! b\bar{b})/BR(h\!\to\! \tau^+\tau^-)$ can discriminate
between SUSY and non-SUSY Higgs models. Assuming a $\pm 5.4$\% measurement of 
this ratio to be made at a 500~GeV ILC, one is sensitive to the SUSY nature of 
$h$ for $M_A$ values up to 1.8~TeV\cite{penaranda}, see Fig.\ref{figguasch}.

\begin{figure}[h]
\begin{center}
\epsfig{figure=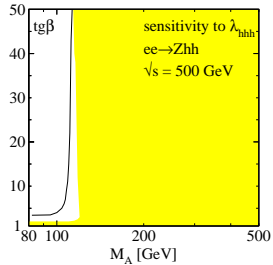,width=3.5cm,clip=}
\end{center}
\caption{Sensitivity to $\lambda_{hhh}$ in $e^+e^-\to Zhh$.}
\label{sens}
\vspace*{-0.3cm}
\end{figure}

There are 6 CP invariant neutral trilinear Higgs self-couplings in the MSSM. 
They are accessible in $WW/ZZ$ fusion into Higgs pairs, 
double Higgs-strahlung and triple Higgs production\cite{3lam,osland}. 
All self-couplings can be determined from these cross sections up to
discrete ambiguities provided they are large enough. Demanding the cross 
sections to exceed 0.01 fb and the effect of a non-zero Higgs self-interaction 
to be larger than 2 st.dev. for $\int{\cal L}=2$~ab$^{-1}$ the coupling among 
three light Higgs bosons is accessible in large ranges of the 
$M_A$-$\tan\beta$ parameter space, see Fig.\ref{sens}\cite{3lam}, other 
couplings are accessible through Higgs cascade decay channels.

In the past a plethora of further extensions of the SM beyond the MSSM 
has been proposed, as {\it e.g.}~the CP violating
MSSM\cite{cp}, the next-to-minimal SUSY extension (NMSSM)\cite{nmssm}, Little
Higgs models\cite{littleh} etc. The precision measurements of couplings 
achievable at a future ILC will help to discriminate and constrain the 
various models. 

\vspace*{-0.3cm}
\section{Summary}
\vspace*{-0.2cm}

The future ILC\cite{itrp} will provide us with precision 
measurements in the electroweak sector of the SM and SUSY
extensions which will be mandatory to understand the Higgs mechanism
in all its essential aspects and thus to understand the mechanism of mass 
generation and EWSB. The high precision on the measured observables 
allows for a sensitivity to new physics at high scales beyond the 
direct reach of the collider itself but also at the LHC.

\vspace*{-0.3cm}
\section*{Acknowledgements}
\vspace*{-0.2cm}

I want to thank A.~Brandenburg, K.~Desch, W.~Kilian, M.~Spira, T.~Teubner and 
P.~M.~Zerwas for helpful discussions and DESY/Hamburg for
the kind hospitality during the preparation of this talk. 


\end{document}